\documentclass[aps,prl,twocolumn,superscriptaddress,showpacs]{revtex4}
\usepackage{amsmath,mathrsfs}
\usepackage{graphicx}
\usepackage{color}

\newcommand{\bi}[1]{\ensuremath{\boldsymbol{#1}}}

\newcommand{\im}{{\rm i}}
\newcommand{\ve}{\varepsilon}
\newcommand{\RH}{R_{\rm H}}

\begin{document}


\title{
Interband Effects of Magnetic Field on Hall Effects for Dirac Electrons in Bismuth
}


\author{Yuki Fuseya}
\email{fuseya@hosi.phys.s.u-tokyo.ac.jp}
\author{Masao Ogata}
\affiliation{Department of Physics, University of Tokyo, Hongo, Bunkyo-ku, Tokyo 113-0033, Japan}
\author{Hidetoshi Fukuyama}
\affiliation{Department of Applied Physics, Faculty of Science, Tokyo University of Science, 1-3 Kagurazaka, Shinjuku-ku, Tokyo 162-8601, Japan}


\date{\today}

\begin{abstract}
Hall effects are investigated for three-dimensional Dirac electrons as a model of bismuth alloys.
%
%
It is found that there is unconventional contributions to the Hall conductivity $\sigma_{xy}$ generated by the interband effects of a magnetic field.
This phenomena is remarkable near the band-edge. 
The Hall coefficient exhibits two unexpected properties; a sharp peak at around the band-edges, and a drastic sign change in the band gap region.
Implications of the present results to bismuth alloys are discussed.

\end{abstract}



\maketitle



	Electrons in a periodic potential can be described as Bloch band theory, which has been proved very useful for many purposes.
	In this theory, it is natural to expect that the effects of an external magnetic field $H$ are described by the effective Hamiltonian $\mathscr{H}_{\rm eff}=E_n (\bi{k}+e\bi{A}/c)$ except for the Zeeman effect, where the $n$-th Bloch band is characterized by an energy function $E_n (\bi{k})$ for $H=0$.
	This simple procedure has been proved to be very useful for interpreting many phenomena.
	Nevertheless, this Bloch band picture is insufficient in principle\cite{Peierlstext,Kubo};
	electrons under a magnetic field are not confined in a single Bloch band, but undergo complex interband oscillations.
	These effects will be referred to as interband effects of a magnetic field, which are not contained in the Bloch band picture.
	Although the interband effect is a fundamental problem even in the non-interacting electron systems, it has been studied only in limited cases\cite{Fukuyama06}, i.e., the orbital magnetism, where the interband effect plays a crucial role\cite{Fukuyama70}.
	In this letter, we study the interband effects on the Hall effects having Bi and its alloys in mind as a typical experimental situation.
	%
	%
	%
	%
	
\begin{figure}[bh]
	\begin{center}
	\includegraphics[height=3.5cm]{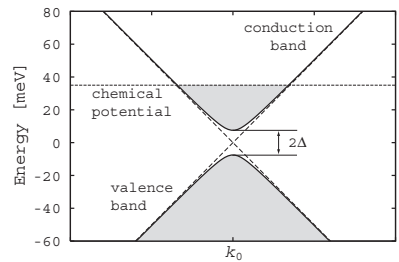}
	\end{center}
	\caption{Band structure of pure bismuth near $L$-point in the Brillouin zone. 
	The chemical potential is $\mu\sim 27$meV from the band-edge of the conduction band, and the band gap is $2\Delta \sim 14$meV\cite{Liu}.}
	\label{2band}
\end{figure}

	Bismuth and its alloys have been famous for their large diamagnetism for a long time.
	Especially they take largest values when the chemical potential is located in the region where the density of states (DOS) vanishes\cite{Wehrli}.
	This can never be understood in the Bloch band picture, i.e., the Landau-Peierls formula\cite{Peierls}, which indicates that the orbital susceptibility $\chi$ is proportional to the DOS.
	This mystery has been solved by taking account of the interband effect\cite{Fukuyama70}.
	In particular, Bi is a semimetal and its band structure has a special form, i.e., a narrow-gap Dirac electron systems (Fig. \ref{2band}).
	This is the reason why the interband effects on orbital magnetism are prominent in Bi.
	%
	%
	%
	%
	%

	%
	The Hall conductivity, $\sigma_{xy}$, is more complex than $\chi$.
	The orbital susceptibility originates from the permanent orbital current, which is non-dissipative.
	On the other hand, $\sigma_{xy}$ in weak magnetic field is essentially dissipative.
	In the following, we study $\sigma_{xy}$ in a three-dimensional Dirac system to show that $\sigma_{xy}$ is actually related to the diamagnetic current.
	Our results give some clues to a long-standing problem about how electrons contribute to both dissipative and non-dissipative currents in the presence of weak magnetic field.
	%
	%
	%
	%
	%
	Recently, some related studies have tackled this problem inspired by newly discovered Dirac electron materials, such as graphene and $\alpha$-ET$_2$I$_3$, which are two-dimensional gapless Dirac electron systems\cite{Fukuyama07,Kobayashi,Nakamura}.
	Although the anomalous behaviors of $\chi$ and the Hall coefficient $\RH$ have been reported, their relationship has not yet been clarified.
	Here the relationship between $\chi$ and $\RH$ in weak magnetic fields resulting from interband effects will be investigated, in detail, in the presence of the finite band gap $2\Delta$.
	%
	%
	%
	%
	%
	%
	
	%
	Several experiments on $\RH$ of Bi have suggested that $\RH$ is quite anomalous, e.g., rapid sign changes\cite{Jain,Brandt69,Abeles,Lerner}, and 3D fractional quantum Hall effect\cite{Behnia}.
	In spite of these fascinating behaviors, there has been no theoretical analysis especially in the weak field limit.
	%
	%
	%
	%
	%
	For a two band system with a small band gap $2\Delta$, such as Bi, it is puzzling how $\RH$ behaves in the band gap region $|\mu|<\Delta$.
	Here, $\mu$ the chemical potential, and the origin of the energy is taken at the center of the band gap.
	In the free electron system, $\RH$ is proportional to $1/n_{\rm e}ec$, $n_{\rm e}$ being the density of electrons.
	Thus it should diverge for $|\mu|<\Delta$, since $n_{\rm e}=0$.
	More precisely, if we start from the general expression, $\RH=\sigma_{xy}/H\sigma_{xx}\sigma_{yy}$, valid even in the multi-band cases, we obtain $\RH\to0/0$ for $|\mu|<\Delta$.
	(Here $\sigma_{\mu\nu}$ is the conductivity tensor.)
	Namely, it is not trivial whether $\RH$ converges or diverges and then the property of $\RH$ in the band gap region is quite puzzling.

	In this Letter, we calculate $\sigma_{xx}$, $\sigma_{xy}$ and $\RH$ in the three-dimensional Dirac electron systems on the basis of gauge invariant Kubo formula in the Luttinger-Kohn representation\cite{LK,Fukuyama69,Fukuyama71}.
	We find that the interband effect induces unconventional contribution to $\sigma_{xy}$ near the band-edge (i.e., $|\mu|\sim \Delta$), and $\RH$ exhibits unexpected features; 
	a sharp peak structure at around $\mu=\pm\Delta$ and a drastic sign change through $\mu=0$.
	We discuss implications of the present results to Bi and its alloys.

	%
	The fact that low-energy properties of Bi can be described by Dirac electrons was first introduced by Cohen and Blount\cite{Cohen}, and then by Wolff in an elegant way\cite{Wolff}.
	For pure Bi, electrons locate near the $L$-point in the Brillouin zone and holes near the $T$-point.
	Previous studies concentrated on the $L$-point electrons, since their contribution is dominant due to much larger velocity of electrons than that of holes at the $T$-point.
	Based on this model, anomalous behaviors of Bi, such as a large g-factor\cite{Cohen}, magneto-optics\cite{Wolff}, large diamagnetism\cite{Fukuyama70} have been clarified.
	%
	%
	Electrons around $L$-points (also holes around $T$-points) are essentially described by the tilted Dirac equations, i.e., Dirac equations with anisotropic velocities\cite{Wolff,Kobayashi}.
	However in order to avoid complexities, which are not essential to present studies, the velocity of electrons, $v$, are assumed to be isotropic for simplicity.
	Then the Hamiltonian for electrons and holes is given by
\begin{align}
	&\mathscr{H} =
	\Delta \beta + \im v  \sum_\mu k_\mu \beta \alpha_\mu ,
\end{align}
	%
	%
	where $\beta$ and $\alpha_i$ are the $4\times4$ matrices that appear in the Dirac theory
\begin{align}
	\alpha_i
	=\left(
	\begin{array}{cc}
	0 & \sigma_i\\
	\sigma_i & 0 \\
	\end{array}
	\right) ,
	\,\,
	\beta
	=\left(
	\begin{array}{cccc}
	1 & 0 & 0 & 0\\
	0 & 1 & 0 & 0\\
	0 & 0 & -1 & 0\\
	0 & 0 & 0 & -1\\
	\end{array}
	\right) ,
\end{align}
	with $\sigma_i$'s being the Pauli spin matrices.
	%
	%
	%
	This ``relativistic" Hamiltonian already includes spin-orbit interactions\cite{Wolff,Fukuyama70}, which is very strong in Bi contrary to graphene or $\alpha$-ET$_2$I$_3$.

	It should be emphasized here that this Hamiltonian is not in the Bloch representation but in the Luttinger-Kohn (LK) representation\cite{LK}, in which the wave function, $\psi_{n\bi{k}}$, is expressed in the form $\psi_{n\bi{k}}=u_{n\bi{k}_0}(\bi{r})e^{{\rm i}\bi{k}\cdot\bi{r}}$, $u_{n\bi{k}_0}$ being the periodic part of the Bloch function at a wave number $\bi{k}_0$, e.g., the $L$-point for electrons or $T$-point for holes.
	The wave function under the magnetic field is correctly represented only in this representation\cite{Fukuyama69}.
	Moreover, in the LK representation, we can obtain straightforwardly the gauge-invariant results for various correlation functions\cite{Fukuyama69,Fukuyama71}.
	Note that this representation is exact and related to the ordinary Bloch wave function by a unitary transformation.
	%
	%
	%

	%
	%
	The results of the conductivity, $\sigma_{xx}$, and the Hall conductivity, $\sigma_{xy}$, (for detail calculations, see Ref. \cite{Fukuyama69}) are summarized as follows:
	\begin{align}
		\sigma_{xx}
		&=
		-\frac{e^2}{\pi^3v}
		\int_{-\infty}^\infty \!\!\!\! d\ve
		f' (\ve-\mu)
		\int_0^\infty \!\!\!\! dX
		\left[
		F_1 \left(\ve, X\right)
		-F_2 \left(\ve, X\right)
		\right],
		\\
		F_1 \left(\ve, X\right)
		&=\frac{X^2(\ve^2+\Gamma^2-\frac{1}{3}X^2-\Delta^2)}
		{\left\{ (\ve+\im\Gamma)^2-X^2-\Delta^2\right\}
		\left\{ (\ve-\im\Gamma)^2-X^2-\Delta^2\right\}},
		\\
		F_2 \left(\ve, X\right)
		&=\frac{X^2\left\{ (\ve+\im\Gamma)^2-\frac{1}{3}X^2-\Delta^2 
		\right\}}
		{2\left\{(\ve+\im\Gamma)^2-X^2-\Delta^2 \right\}^2}
		+{\rm c.c.},
	\end{align}
	\begin{align}
		\sigma_{xy}
		&=\frac{e^3vH}{12\pi^2c}
		\int_{-\infty}^\infty \!\!d\ve
		\left[
		F_3 (\ve)
		f(\ve-\mu)
		%
		+F_4 (\ve)
		f'(\ve-\mu)
		\right]{\rm sgn}(\ve),
		\label{cxy}
		\\
		F_3 (\ve)
		&=\frac{\ve+\im\Gamma}
		{\left\{ (\ve+\im \Gamma )^2-\Delta^2\right\}^{3/2}}
		+{\rm c.c.},
		\\
		F_4 (\ve)
		&=
		\frac{-2\Gamma^4-\Gamma^2\Delta^2+(\Delta^2-\ve^2)^2
		+2\im \Gamma^3\ve -\im \Gamma\mu(\Delta^2-\ve^2)}
		{2\Gamma^2\ve^2\sqrt{\ve^2-\Gamma^2-\Delta^2+2\im \Gamma\ve}}
		\nonumber\\&
		+{\rm c.c.},
\end{align}
	%
	where $f(\ve)$ is the Fermi distribution function.
	Here we have introduced a finite damping, $\Gamma$, for electrons as in Ref. \cite{Fukuyama07} to represent the effects of  impurity scattering present in actual materials.
	(We assume $\Gamma$ to be constant in order to make our argument as simple and transparent as possible, although Shon and Ando have indicated that $\Gamma$ somewhat depends on energy and momentum\cite{Shon}.)
	%
	%
\begin{figure}[h]
\begin{center}
\includegraphics[width=6.5cm]{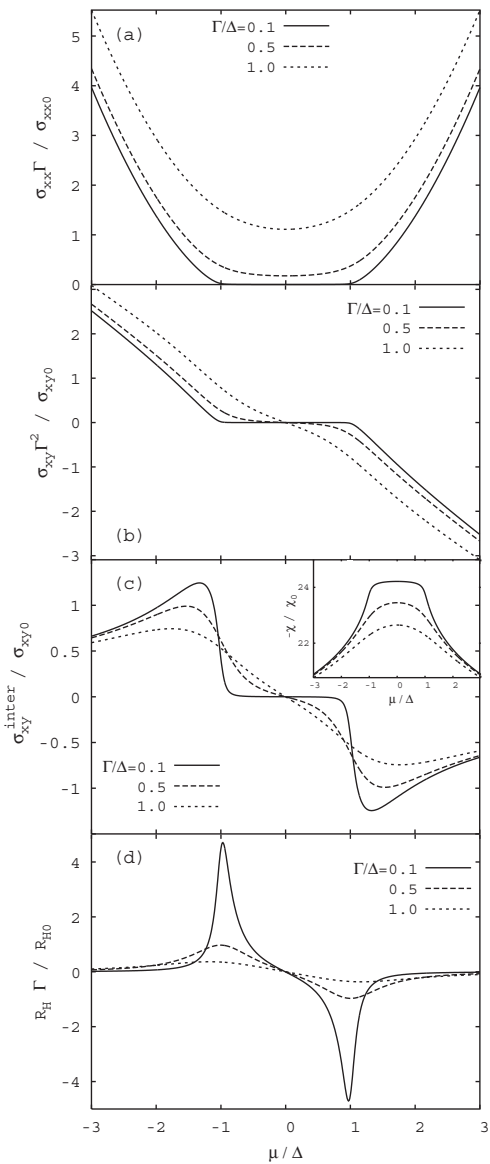}
\end{center}
\caption{The chemical-potential dependences of (a) the conductivity $\sigma_{xx}$, (b) the Hall conductivity $\sigma_{xy}$, (c) the interband contribution to $\sigma_{xy}^{\rm inter}$, and (d) the Hall coefficient $\RH$ at $T=0$.
The inset of (c) is that of the orbital susceptibility $\chi$.
}
\label{all}
\end{figure}
	%
	%
	The chemical potential dependence of $\sigma_{xx}$ and $\sigma_{xy}$ at $T=0$ are shown in Fig. \ref{all} (a) and (b), respectively.
	The normalization factors are taken to be $\sigma_{xx0}=e^2/\pi^2v$ and $\sigma_{xy0}=e^3v/12\pi^2c$.
	Away from the band-edge, $|\mu| \gg \Delta$, $\sigma_{xx}\propto \mu^2$ and $\sigma_{xy}\propto -\mu$, which are consistent with the Bloch band picture.
	In the band gap region, $|\mu|<\Delta$, both conductivities have only small values due to $\Gamma$.
	These non-zero values produce an anomalous $\RH$ as shown below.

	%
	The Hall coefficient defined by $\RH=\sigma_{xy}/\sigma_{xx}^2H$ is shown in Fig. \ref{all} (d) where the normalization is taken to be $R_{{\rm H}0} = \sigma_{xy0}/\sigma_{xx0}^2 H$.
	Away from the band-edge, $\RH \propto -\mu^{-3}$ as is expected from the properties of $\sigma_{xx}$ and $\sigma_{xy}$.
	This is consistent with the Bloch band picture since carrier density is proportional to $|\mu^3|$ in three dimension.
	For $|\mu|\leq \Delta$, on the other hand, $\RH$ exhibits unexpected features; a peak structure at $\mu\simeq \pm\Delta$ and a rapid sign-change through $\mu=0$.
	Note that there is no carriers at $T=0$ of $|\mu|< \Delta$ in clean systems. 
	It is rather surprising that $\RH$ does not diverge but has small values (esp. $\RH=0$ at $\mu=0$) for such small carrier density.
	The very particular feature, $\RH=0$ at $\mu=0$, is similar to the case of gapless Dirac electrons in two-dimension\cite{Fukuyama07,Kobayashi}.
	%
	%
	%

	Now we study the interband effects of magnetic field on $\sigma_{xy}$.
	We extract the interband contribution, $\sigma_{xy}^{\rm inter}$, by subtracting the intraband contribution, $\sigma_{xy}^{\rm intra}$, from $\sigma_{xy}$, namely,
\begin{align}
	\sigma_{xy}^{\rm inter} = \sigma_{xy} -\sigma_{xy}^{\rm intra}.
\end{align}
	Here $\sigma_{xy}^{\rm intra}$ is the Hall conductivity calculated within the intraband approximation, i.e., the Bloch band picture (cf. \cite{Fukuyama69}), which is given by
\begin{align}
	\sigma_{xy}^{\rm intra}
	&=-\frac{e^3vH}{6\pi^3c}\sum_{n=\pm}
	\int_{-\infty}^{\infty}\!\! d\ve 
	f'(\ve -\mu)
	\int_0^\infty \!\! dX
	\frac{n X^4 }{\left[ E_n (X) \right]^3}
	\nonumber\\
	&\times
	\frac{4\Gamma^3}{3\left[ (\ve-E_n(X))^2+\Gamma^2\right]^3},
	\label{cxy_intra}
\end{align}
	where $E_\pm(X)=\pm\sqrt{X^2+\Delta^2}$.
	In spite of the apparently different expression between Eqs. (\ref{cxy}) and (\ref{cxy_intra}), they agree with each other except for the band-edge region;
	$\sigma_{xy}^{\rm intra}$ has only small value for $|\mu|< \Delta$, and increases as $\sigma_{xy}^{\rm intra}\sim -\mu$ for $|\mu|>\Delta$.
	%

	The obtained interband contribution $\sigma_{xy}^{\rm inter}$ is shown in Fig. \ref{all} (c).
	The remarkable property is that $\sigma_{xy}^{\rm inter}$ takes the largest value at the band-edge, and becomes smaller away from the band-edge, contrary to $\sigma_{xy}^{\rm intra}$.
	Furthermore, $\sigma_{xy}^{\rm inter}$ does not depend on $\Gamma$ so much, while $\sigma_{xy}^{\rm intra}$ does.
	(Note that the vertical axis of Fig. \ref{all} (b) includes a factor $\Gamma^2$.).
	This indicates that $\sigma_{xy}^{\rm inter}$ has a different nature from $\sigma_{xy}^{\rm intra}$.
	As seen in the inset of Fig. \ref{all} (c), which shows orbital susceptibility $\chi$ calculated in the gauge-invariant formula\cite{Fukuyama71}, $\sigma_{xy}^{\rm inter}$ appears to be associated with diamagnetic currents.

	In the following, we discuss the physical origin of the interband contribution $\sigma_{xy}^{\rm inter}$ in relation to the orbital susceptibility.
	In the band gap region, electrons in a magnetic field circulate locally and make diamagnetic current.
	There are no electrons going through the crystal because of the band-gap.
	In the band-edge region, in contrast, the electrons start to be transferred from local diamagnetic orbitals to the neighboring orbitals due to some scattering.
	These processes will generate contributions to the Hall conductivity: $\sigma_{xy}^{\rm inter}$.
	As is known from the behavior of $\chi$ (Fig. \ref{all} (c)), the diamagnetic current has largest values for $|\mu|\lesssim\Delta$, and decreases away from the band-edge $|\mu|\gg \Delta$.
	Correspondingly, $|\sigma_{xy}^{\rm inter}|$ actually decreases away from the band-edge.
	%
	%
	%
	%
	Note that in the limit of $\Delta/\Gamma \to 0$, $\sigma_{xy}^{\rm inter}$ is given only by $F_3$ term of Eq. (\ref{cxy}), where the factor $f(\ve -\mu)$ generates the contributions of bands below $\mu$.

\begin{figure}[tbh]
\begin{center}
\includegraphics[width=7cm]{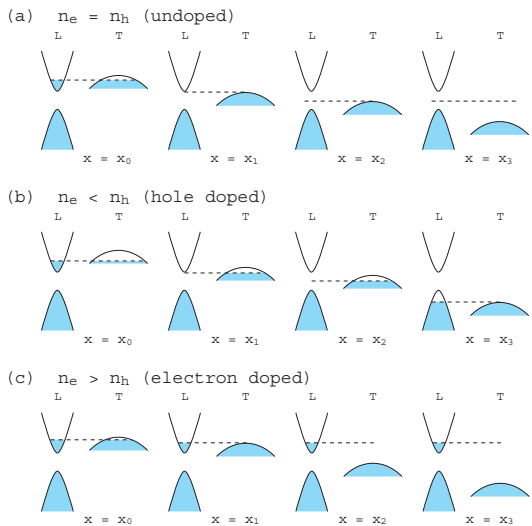}
\end{center}
\caption{Schematic band structure around $L$- and $P$-points of Bi and its alloys for (a) undoped ($n_{\rm e}=n_{\rm h}$), (b) hole doped ($n_{\rm e}<n_{\rm h}$), and (c) electron doped ($n_{\rm e}>n_{\rm h}$) case. The dashed lines indicate the level of the chemical potential.
}
\label{LTnp}
\end{figure}
	%
	Finally, let us discuss the implications of the present results to Bi.
	Some measurements of $\RH$ in Bi have already exhibited the peak structures and the rapid sign changes with respect to $\mu$\cite{Jain,Brandt69}.
	They are, however, not consistent with each other.
	This would be due to the purity of samples or the inhomogeneity of either chemical or external pressure.
	For further investigations, we need some guidelines.
	%
	%
	%
	It will be rather difficult to find the peak of $\RH$ at $\mu\simeq \pm\Delta$ when the peak is very sharp.
	So we propose a simultaneous measurement of $\chi$ together with that of $\RH$ to observe the band-edge property in a more transparent way.
	%
	%
	%
	%
	A clear kink of $\chi$ at $\mu\simeq \pm\Delta$ has been actually observed experimentally\cite{Wehrli}.
	Thus, the kink in $\chi$ can be used to identify the band-edge, and then we can expect the peak of $\RH$ in the same region.
	%
	%
	%
	%
	%
	%
	
	When Bi alloys are used, it will be important to consider the contributions of both electrons and holes.
	%
	%
	%
	%
	%
	%
	%
	%
	The band structure around $L$- and $T$-points are illustrated in Fig. \ref{LTnp} for (a) undoped Bi ($n_{\rm e}=n_{\rm h}$), (b) hole doped Bi ($n_{\rm e}<n_{\rm h}$), and electron doped Bi ($n_{\rm e}>n_{\rm h}$). 
	Here, $x$ denotes some controlling parameter which changes the relative position of electron- and hole-band, for example, external pressure or alloy concentration.
	For undoped case (Fig. \ref{LTnp} (a)), a clear band-edge property --- a peak in $\RH$ and a kink in $\chi$ --- can be seen, since $n_{\rm e}$ and $n_{\rm h}$ vanish simultaneously at $x=x_1$.
	This band-edge property will be clearer when $H$ is applied perpendicular to $z$ (trigonal axis), since the contribution of electrons is dominant\cite{Fukuyama70}.
	In this case, the sign change in $\RH$ will not be observed, since $\mu$ is always positive for electrons.
	For the hole doped case (Fig. \ref{LTnp} (b)), on the other hand, $\mu$ of electrons can be negative for $x>x_2$, so that the sign change in $\RH$ is possible.
	However, the band-edge property in $\RH$ will not be so clear due to the finite contribution from the holes at $T$-point, even though it will be clear in $\chi$.
	For the electron doped case (Fig. \ref{LTnp} (c)), neither the band-edge effect nor the sign change will be hardly seen.

	We have studied the Hall effects of the Dirac electron systems with special reference to Bi on the basis of the Kubo formula in the Luttinger-Kohn representation keeping gauge invariance.
	It is clarified that the interband effect of a magnetic field generates the novel contribution to $\sigma_{xy}$.
	This contribution due to interband effect, $\sigma_{xy}^{\rm inter}$, is remarkable near the band-edge. 
	This property is quite different from the conventional Hall conductivity which originates from a single Bloch band.
	%
	%
	The Hall coefficient exhibits two unexpected properties; a sharp peak at $\mu= \pm\Delta$ and a rapid sign-change through $\mu=0$.
	We can expect a clear structure of $\RH$ even in the band gap region $|\mu|\le \Delta$.
	%
	%
	In order to detect the band-edge effect clearly, we have proposed simultaneous measurements of $\RH$ together with $\chi$.
	We have predicted that the band-edge property and the rapid sign-change can be seen for the undoped Bi alloys ($n_{\rm e}=n_{\rm h}$) and for the hole doped Bi alloys ($n_{\rm e}<n_{\rm h}$), respectively.

	
	We thank A. Kobayashi for helpful discussions.
	The present work was financially supported by Grants-in-Aid for Scientific Research on Priority Areas of Molecular Conductors (No. 15073210) from the Ministry of Education, Culture, Sports, Science and Technology, Japan.
	Y.F. is supported by JSPS Research Fellowships for Young Scientists.

\bibliography{basename of .bib file}

\end{document}